\documentclass{aa}

 \usepackage{savesym}       
    \savesymbol{tablenum}
    \usepackage{siunitx}
        \restoresymbol{SIX}{tablenum}   
 \usepackage{wrapfig} 
 \usepackage{hyperref}
 \usepackage{subfigure}
\usepackage{xcolor}
\usepackage{amsmath} 

\begin{document}

\title{Analysing the highly irregular boundaries of solar pores}

\author{
T.~J.~Duckenfield\inst{1}
\and
D.~B.~Jess\inst{1,2}\corrauth{d.jess@qub.ac.uk}
\and
S.~Jafarzadeh\inst{1}
\and
L.~A.~C.~A.~Schiavo\inst{3}
\and
S.~S.~A.~Silva\inst{4}
\and
S.~D.~T.~Grant\inst{1}
}

\institute{
Astrophysics Research Centre,
School of Mathematics and Physics,
Queen's University Belfast,
Belfast, BT7 1NN, UK
\and
Department of Physics and Astronomy, California State University Northridge, Northridge, CA 91330, USA
\and
Northumbria University,
Newcastle upon Tyne,
NE1 8ST, UK
\and
Plasma Dynamics Group,
School of Electrical and Electronic Engineering,
The University of Sheffield,
Sheffield, UK
}
\abstract
{Solar pores possess irregular and evolving boundaries that are often far from the ideal circular flux tubes assumed in many magnetohydrodynamic (MHD) oscillation models.}
{To tackle this irregularity in a consistent way for wave analysis, we introduce a framework that employs the convex hull of the pore boundary --- derived from intensity minimum error thresholding --- as the domain to perform further analysis. Using the modal assurance criterion, we find the range of pore shapes for which this approximation is valid.}
{We demonstrate the usefulness of this framework by applying it to multi-height, high-cadence observations (4170\,\si{\angstrom} continuum, G-band, Na~\textsc{i}, and Ca~\textsc{ii}~K) of a solar pore, and apply Proper Orthogonal Decomposition  of the convex hull to determine wave modes.}
{The fundamental sausage ($m=1$) and kink ($m=2$) mode is found to remain reliable, while higher-order fluting modes ($m\ge3$) are strongly degraded by small-scale boundary irregularity. As expected, sausage-like modes dominate the variance at all heights and exhibit a systematic upward shift in frequency, consistent with freely propagating compressive waves. In contrast, the kink-like motions appear weak, confined to a persistent low-frequency peak, and most plausibly interpreted as a forced response to granular buffeting rather than a propagating mode.}
{Together, these results establish a practical methodology for boundary-mode analysis in real, highly structured pores and provide new constraints on the nature and height evolution of MHD waves in the lower solar atmosphere.}


\maketitle
\nolinenumbers
\section{Introduction} \label{sec:intro}
Solar pores occupy a curious middle ground in terms of magnetic structures on the Sun: seemingly simpler than sunspots yet sufficiently magnetised to act as efficient waveguides for a rich spectrum of magnetohydrodynamic (MHD) oscillations \citep{Morton2011_pore, Grant2015, Keys2018, Gilchrist-Millar2020,Jess2023_obsreview}. 
Their smaller size means they are more dynamic and responsive to external forces, unlike sunspots whose stronger magnetic fields make the umbral-penumbral boundary more rigid.
Indeed recent studies have revealed that pores rarely oscillate as coherent, monolithic structures; instead, the lack of phase coherence around their perimeters implies significant small-scale corrugation \citep[e.g.,][]{Grant2022}.
Solar pores are buffeted by granular flows which drive continual forcing of the pore boundary, shaping both their long-term evolution and the short-timescale oscillations, such that even small scale changes of plasma motions are conducive to converting pores into sunspots \citep{Kamlah2023}.

High-cadence observations over the past decade have shown that pores support both compressive and transverse wave modes that propagate upward through the lower atmosphere, often with measurable damping as they traverse the steep stratification \citep{Grant2015,Jess2012_simulatedupwards, Freij2016, Keys2018, Freij2016, Gilchrist-Millar2020}.
The steep gradients in magnetic field strength, density, and acoustic cut-off frequency imply that pores act as height-dependent filters for particular MHD modes. 
For example, it was shown in \citet{Miriyala2025} that the temperature structure of the lower atmosphere naturally filters upward-propagating acoustic waves, truncating the power spectra for low frequencies. The subsequent mode conversions -- i.e., fast-to-fast and fast-to-Alfv\'en -- are linear, so the filtered power spectrum of the $p$-modes is imparted on the resultant transformed wave modes.
Such wave modes can transport significant energy into the chromosphere, potentially contributing to atmospheric heating \citep{Guevara-Gomez2023}.

Recent results have highlighted the need to disentangle externally driven responses from genuine resonant behaviour in small magnetic flux concentrations. 
For example, \citet{Stangalini2021_Bomega} present the $B$-$\omega$ technique for a small pore, displaying a rich spectrum of MHD modes which do not necessarily match the $p$-mode driving spectrum. 
This directly implies a mixture of driven and resonant oscillatory behaviour.
These studies established pores as natural laboratories for diagnosing wave energetics and for testing the interplay between geometric structure, magnetic field topology, and mode behaviour.

Multi-height observations are extremely powerful for connecting boundary-driven perturbations in the photosphere to their upward-propagating counterparts in the low chromosphere, since different spectral diagnostics sample different atmospheric heights and radiative regimes \citep{Cauzzi2008}.
Nonetheless direct observational constraints on how individual mode families evolve with height remain sparse.

Proper Orthogonal Decomposition (POD) is a data-driven technique designed to extract coherent spatial structures and wave modes from spatio-temporal datasets, ranking them by their contribution to the total signal variance. In solar physics, it was first applied to the detection of waves in the sunspot photosphere \citep{Albidah_2022, Albidah_2023} and, more recently, in photospheric magnetic flux ropes \citep{Alanezy2026}. 
The technique has since been extended to pore boundaries observed with Solar Orbiter/PHI, where it revealed well-structured sausage, kink, and higher-order fluting modes even at relatively modest spatial resolution \citep{Jafarzadeh2024_fluting}. 
POD has also been employed to identify global compressible oscillations in the outer solar corona driven by a coronal mass ejection \citep{Silva_2025}, as well as to characterise magnetoacoustic and Alfvén waves in numerical simulations of magnetic reconnection \citep{schiavo2026waves}. 
In this work, we adopt the implementation described in \citet{Jafarzadeh2024_fluting}, which is available as part of the WaLSAtools library \citep{Jafarzadeh2025}.
At the sub-arcsecond resolution of modern ground-based facilities, however, pore boundaries become increasingly fragmented and finely structured. 
In such circumstances the boundary may no longer behave as a smooth, coherent wave-supporting surface, as indicated by the phase-fragmented behaviour reported by \citet{Grant2022}. 
This raises an important question for high-resolution observations: to what extent can we still extract meaningful surface-mode signatures from boundaries that are dynamically corrugated?

In this paper we address this question directly by analysing multi-height imaging observations of a typical photospheric pore and applying POD to its boundary dynamics after carefully regularising the geometry. 
Our aim is twofold: first, to determine which deformation modes remain physically robust in the presence of realistic boundary complexity; and second, to characterise how these modes evolve with height as they propagate from the deep photosphere into the low chromosphere. 

\section{Observations}
\label{sec:observations}
Utilising the same high-cadence dataset analysed by \citet{Grant2015}, an irregularly shaped solar pore is observed across four photospheric and chromospheric bandpasses: the 4170\si{\angstrom} continuum, G-band, Na\,{\sc{i}}, and Ca\,{\sc{ii}}\,K. 
These channels sample progressively higher layers of the lower solar atmosphere, providing a multi-height view of the evolving pore boundary. 
All datasets were co-aligned and interpolated onto a common grid with a plate scale of \qty{0.12}{\arcsec} per pixel and an effective cadence of \qty{2.11}{\second}. 
The observations were acquired under excellent seeing conditions for a continuous \qty{35}{minute} interval (19:27–20:02~UT on 2013 March~6) using the Rapid Oscillations in the Solar Atmosphere \citep[ROSA;][]{Jess2010_ROSA} instrument at the Dunn Solar Telescope. 
The pore, located at helioprojective coordinates (122'', -155''), has an average diamater of approximately \qty{3.6}{Mm} and remains largely isolated from nearby magnetic activity throughout the sequence.
\begin{figure}
    \centering
    \includegraphics[angle=90, width=\linewidth]{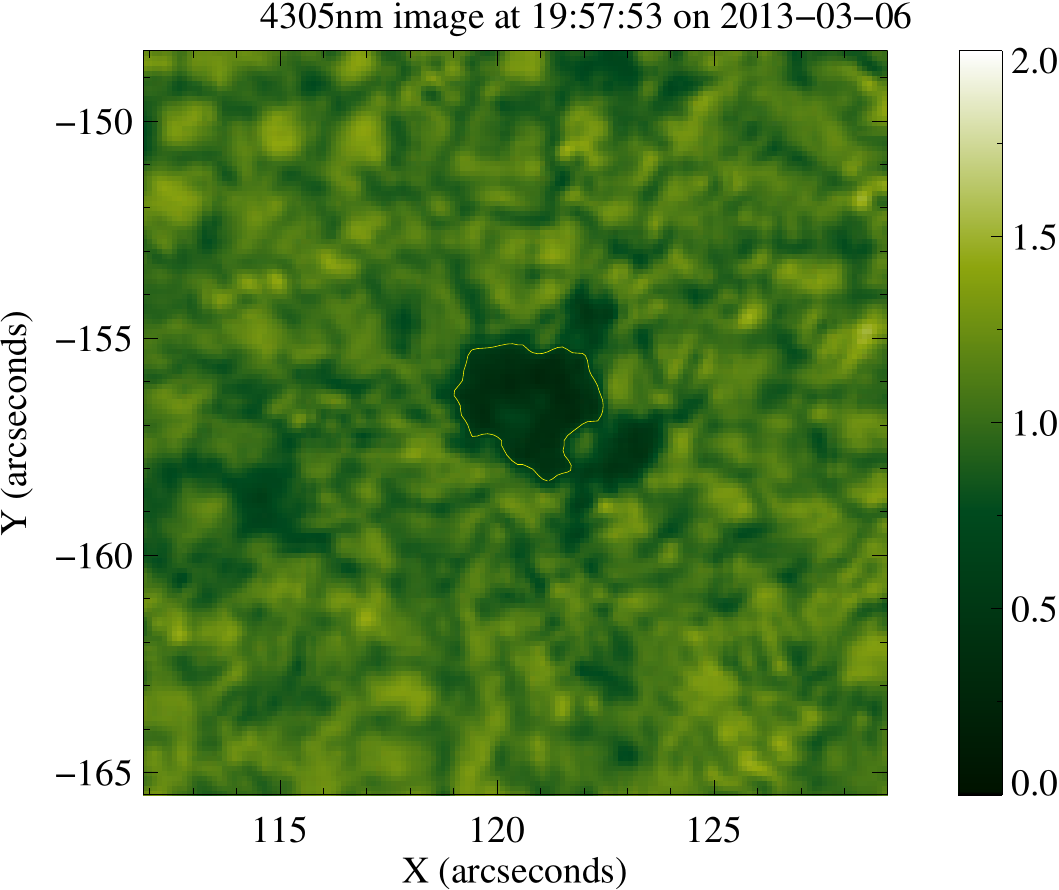}
    \caption{G-band context image of the pore, with the boundary extracted via Minimum Error Thresholding overplotted in yellow. 
    }
    \label{fig:Gband_context}
\end{figure}

\section{Methodology}
\label{sec:method}

\subsection{Boundary extraction}
Given the inherent complexity of solar observations, including the effects of variable seeing conditions and the evolving, non-trivial shape of the pore boundary, the edge of the pore is determined using a data-driven segmentation technique based on Minimum Error Thresholding applied to the intensity histogram in and around the pore for each frame. 
Minimum Error Thresholding (also known as the Kittler-Illingworth algorithm) selects the boundary-defining intensity level by choosing the optimal threshold $T$ that minimises the probability of misclassifying pixels between the pore interior and the surrounding photosphere, 
\begin{equation}
E(T) = P_1(T)\,E_1(T) + P_2(T)\,E_2(T),
\end{equation}
where $P_1$ and $P_2$ are the probabilities of a pixel belonging to the foreground or background, and $E_1$ and $E_2$ are the corresponding misclassification errors determined from their intensity distributions.
The threshold $T$ is chosen such that $E(T)$ is minimised.
The resulting binary mask defines a simply connected pore interior, from which we obtain the boundary as a closed, single-valued contour. 

This frame-adaptive scheme is particularly important when comparing multiple bandpasses with differing contrast and radiative formation properties, since diagnostics formed at different atmospheric heights can exhibit substantially different apparent morphologies and intensity structures \citep{Cauzzi2008}. 
Because MET responds only to the instantaneous bimodality of the local intensity distribution, it naturally tracks seeing fluctuations and avoids tuning a fixed offset or threshold parameter.


\subsection{Convex hull}
At the high spatial resolution of these observations, the pore boundaries are often riddled with small-scale corrugations and deep indentations, supported by the lack of phase coherence \citep{Grant2022}.
This small scale complexity makes it challenging to use the shape directly as a wave-supporting boundary for surface-mode analysis.

To obtain a well-behaved analysis domain that retains the characteristic size and overall footprint of the pore, and allow us to search for waves using techniques such as POD, we replace the extracted boundary with its \textit{convex hull}, defined as the smallest convex polygon enclosing the pore. 
The convex hull is a standard tool for regularising irregular geometries and has already proven useful in solar physics, for example in studies of active region oscillations \citep{Dumbadze2017_pt1}. 
The resulting hull is interpolated to a uniform arc-length parametrisation.

\begin{table*}
\caption{Shape metrics for the pore boundary across four examples, one from each bandpass. 
The final row IsoQ / IsoQ$_{\rm convex}$ shows the ratio of circularities before and after convexification, and acts as a useful proxy.}
\label{tab:shape_metrics}
\centering
\begin{tabular}{lcccc}
\hline\hline
Metric & Continuum & G-band & Na\,\textsc{i} & Ca\,\textsc{ii}\,K \\
\hline
Circularity                & \num{0.425954} & \num{0.727288} & \num{0.204573} & \num{0.639428} \\
Circularity (convex hull)  & \num{0.913037} & \num{0.927810} & \num{0.901362} & \num{0.849733} \\
Eccentricity               & \num{0.356567} & \num{0.208824} & \num{0.288162} & \num{0.645480} \\
Roughness                  & \num{1.53221}  & \num{1.17259}  & \num{2.21094}  & \num{1.25056}  \\
Roughness (convex hull)    & \num{1.04654}  & \num{1.03817}  & \num{1.05330}  & \num{1.08482}  \\
Solidity                   & \num{0.699844} & \num{0.898384} & \num{0.599696} & \num{0.859028} \\
IsoQ / IsoQ$_{\rm convex}$ & \num{0.466525} & \num{0.783877} & \num{0.226959} & \num{0.752505} \\
\hline
\multicolumn{5}{c}{Mode similarity} \\
\hline
MAC$_{11}$ (sausage)       & 0.215 & 0.973 & 0.438 & 0.945 \\ 
MAC$_{22}$ (kink)          & 0.020 & 0.912 & 0.206 & 0.743 \\ 
MAC$_{33}$ (fluting)       & 0.139 & 0.863 & 0.007 & 0.657 \\ 
\hline
\end{tabular}
\end{table*}
A shape's convex hull necessarily increases the enclosed area by an amount that depends on the degree of boundary irregularity.
To quantify the difference between the observed pore shape and its convex hull, we compute several standard shape descriptors. 
The solidity, $A/A_{\mathrm{convex}}$, measures the fraction of the hull area actually filled by the pore and therefore indicates the level of indentation or fragmentation. 
The roughness, $P / (2\sqrt{\pi A})$, compares the measured perimeter $P$ with that of a circle of equal area, capturing the degree of small-scale boundary scalloping. 
The circularity, $4\pi A / P^{2}$, measures the shape's deviation from a perfect circle, and is also known as the isoperimetric quotient.  
The eccentricity, $\sqrt{1-b^{2}/a^{2}}$, where $a$ and $b$ are the semi-major and semi-minor axes of the best-fit ellipse respectively, quantifies the elongation of the pore geometry.
Together, these metrics allow us to track the evolving geometry of the pore and assess when the convex hull remains a faithful surrogate for the true boundary.

\section{Results}
\label{sec:results}
\subsection{Validating the convex-hull approximation}
\label{subsec:validate}
Because convexification alters the boundary geometry, we quantify its effect using the geometric eigenmodes of both the observed pore shape and its convex hull. 
The use of geometric eigenfunctions as a basis for analysing boundary deformations is supported by recent work showing that the spatial structure of MHD modes in small magnetic elements is strongly shaped by the geometry of the confining flux tube itself \citep{Stangalini2022_Nature}. 
In such cases the eigenfunctions of the boundary provide an appropriate modal basis for interpreting the dominant surface oscillations.

Following the numerical approach of \citet{Aldhafeeri2021,Aldhafeeri2022}, we solve for the first five eigenfunctions of the Laplacian defined for each pore geometry. 
These eigenfunctions form the natural deformation basis for the shape and therefore provide an appropriate space in which to evaluate the impact of geometric simplification.
The solver also returns the associated eigenfrequencies of the boundary-defined eigenproblem. 
These frequencies quantify the natural spectrum supported by the instantaneous geometry of the pore, and not the observed oscillatory frequency of the pore itself. 
Nonetheless relative changes in the eigenfrequency spectrum provide a quantitative measure of how modifications to the boundary geometry would alter the set of supported deformation modes.

We compare corresponding eigenmodes using the Modal Assurance Criterion (MAC), defined as the normalised inner product between two mode shapes (labelled $\phi$ and $\psi$): 
\begin{equation}
\mathrm{MAC}_{ij} =
\frac{\left|\,\boldsymbol{\phi}_i^{\mathrm{T}} \boldsymbol{\psi}_j \right|^2}
     {\left( \boldsymbol{\phi}_i^{\mathrm{T}} \boldsymbol{\phi}_i \right)
      \left( \boldsymbol{\psi}_j^{\mathrm{T}} \boldsymbol{\psi}_j \right)}.
\end{equation}
A MAC value close to unity indicates strong similarity, whereas low values signal modal mixing or reordering. 
Therefore the MAC value quantifies the extent to which the corresponding deformation modes are preserved under convexification, independent of their amplitude or phase. 

\begin{figure}
    \centering
    \includegraphics[width=\linewidth]{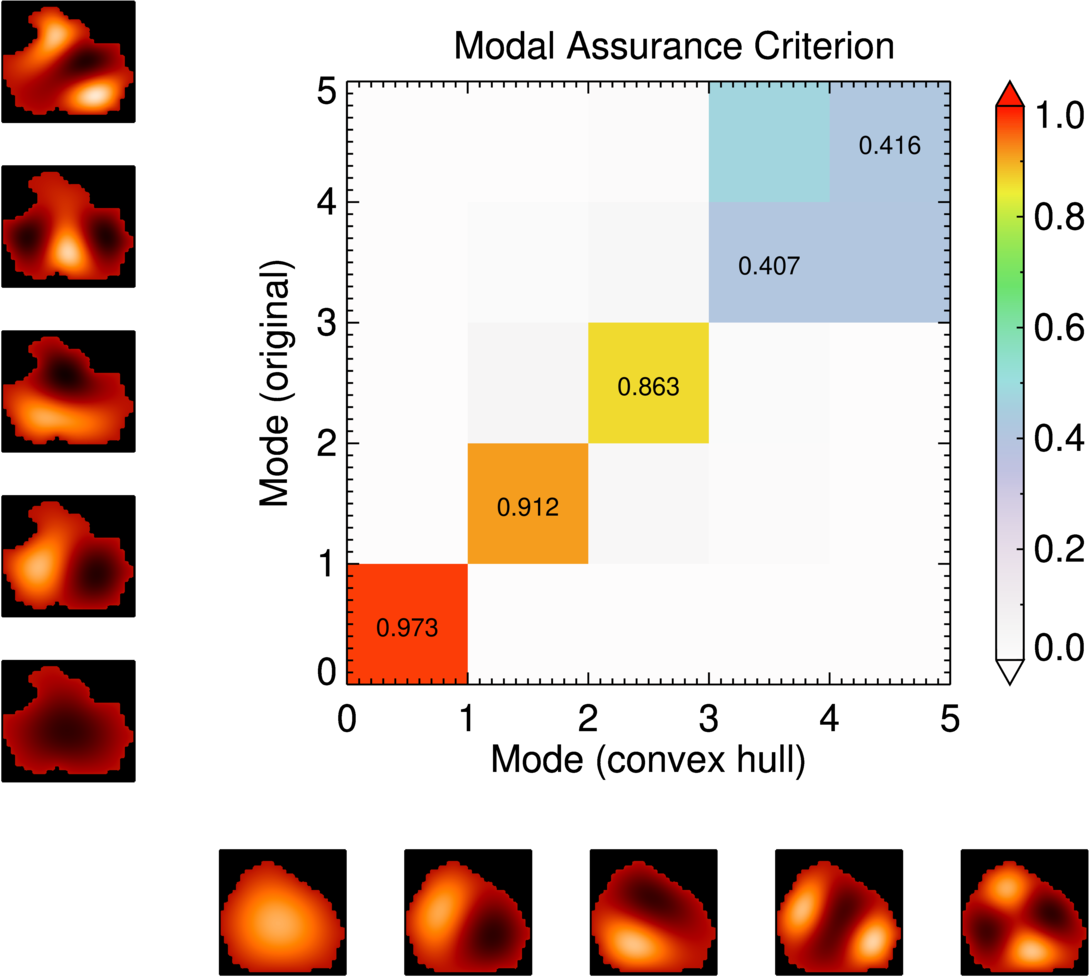}
    \caption{Modal Assurance Criterion matrix between the first 5 eigenmodes of a pore boundary calculated in the G-band, and the first 5 eigenmodes of the pore boundaries convex hull.}
    \label{fig:MAC_example}
\end{figure}

To illustrate how the convex hull can preserve the oscillatory behaviour of a pore boundary, in Figure ~\ref{fig:MAC_example} we compare the first five geometric eigenfunctions for the pore observed in the G-band at one representative instance, 19:57:53, to the first five eigenfunctions of its convex hull. 
We can see from the matrix of Figure~\ref{fig:MAC_example} that the first three modes show strong correspondence, with diagonal MAC values of $0.97$, $0.91$, and $0.86$. 
These high MAC values indicate that the large-scale sausage-, kink-, and the lowest fluting-like deformation modes inherent to the original boundary are retained well by the convex hull.
For the higher-order modes, the agreement is weaker: MAC$_{44}\approx0.41$ and MAC$_{55}\approx0.42$, together with noticeable mixing between these two modes indicated by the nondiagonal elements. 
This reduced similarity reflects the sensitivity of higher-order azimuthal structure to the fine-scale concavities of the real boundary, which are necessarily removed by convexification. 
The increased complexity of the higher-order modes' node locations is clear when looking at the eigenmodes themselves, as is the fact that their orientation shifts going from the original to the convex hull.
Thus, while the convex hull provides a reasonable approximation for the lowest-order, globally coherent deformations, it becomes less representative for higher-order modes that depend more strongly on the detailed boundary shape, consistent with the phase-fragmented behaviour reported in high-resolution observations \citep{Grant2022}.
The geometric eigenfrequencies also shift systematically under convexification. 
Such a frequency shift is generally always expected and will be negative, since the convex hull is always larger in area than the original shape, leading to a reduction in eigenfrequency.
For the G-band example, the first five eigenfrequencies are reduced by between $6 - 10\%$ relative to those of the original boundary, with fractional changes of $-9.7\%$, $-7.0\%$, $-8.4\%$, $-6.5\%$, and $-8.0\%$ for modes 1–5, respectively. 
These shifts reflect the slightly larger and smoother domain defined by the convex hull, which lowers the effective stiffness of the boundary and therefore the associated characteristic frequencies. 
The fact that all modes experience comparable fractional changes indicates that the convex-hull transformation introduces a largely uniform geometric adjustment, consistent with the MAC results that show strong correspondence for the lowest-order modes and more mixing only in the higher-order case.

To determine how broadly the convexification behaviour applies to other pore instances, we repeated the MAC analysis for several pore boundaries spanning the range of shape metrics in Table~\ref{tab:shape_metrics}. 
A consistent pattern emerges. 
Pore boundaries that are intrinsically compact and only weakly corrugated—those with high solidity and low roughness—exhibit strong modal correspondence between the measured boundary and its convex hull. 
In such cases, the sausage and kink modes remain clearly identifiable, and only the higher-order fluting modes show appreciable mixing. 
By contrast, pores with deeply indented or highly fragmented boundaries (roughness $\gg1$, circularity $\ll 1$) do not have a clear mapping of modes from the original shape to their convex hull: their MAC matrices contain no dominant diagonals, and the low-order eigenfunctions of the raw and convexified boundaries bear little resemblance to one another. 
This behaviour reflects the fact that, in highly irregular pores, the geometric eigenmodes of the raw boundary are dominated by short-wavelength concavities; using the convex hull removes these features and therefore yields a fundamentally different modal basis. 
In other words, once the boundary becomes sufficiently rough, any apparent oscillatory `modes' are better interpreted as forced, granular-scale perturbations rather than coherent standing eigenmodes supported by the pore itself.
Eccentricity shows little predictive power of modal correspondence compared to solidity or roughness, and is therefore not seen to be harmful to boundary-mode analysis.
The full MAC matrices and corresponding eigenmodes illustrating these regimes are presented in the Appendix.

A related point concerns boundary analyses performed at lower spatial resolution, such as the POD study of Solar Orbiter/PHI pores by
\citet{Jafarzadeh2024_fluting}. 
Their results highlight the diagnostic power of boundary-based seismology even when the pore outline is only (relatively) coarsely resolved.
In such data the instrumental point-spread function and pixel sampling act to suppress the fine-scale indentations that dominate the high-resolution boundaries examined here. 
In effect, low-resolution measurements introduce a form of convexification, meaning that the retrieved boundary is already close to the compact, weakly corrugated regime in which a stable eigenmode basis exists. 
In contrast, at sub-arcsecond resolution (as is the case here) pore boundaries almost always display intrusions and fine-scale fracturing, which we have shown can overwhelm the coherent low-order deformations. 
For this reason, taking the convex hull of the pore is a sensible way to isolate the large-scale modes that boundary seismology is intended to probe. 

In summary, when the boundary is compact and only weakly corrugated -- typically reflected in high solidity, moderate circularity, and roughness values close to unity -- the convex hull preserves the coherent low-order deformation modes associated with sausage and kink behaviour. 
Once the boundary develops deep concavities or granular-scale fragmentation, however, the notion of a stable eigenmode basis breaks down: in this regime the raw boundary is dominated by short-wavelength, forced distortions rather than by genuine standing modes, and the convex hull simply reveals that no robust large-scale eigenmodes exist to preserve. 
These criteria therefore mark the practical range over which using the convex hull remains meaningful for boundary-mode analysis.

\begin{figure*}
    \centering
    \includegraphics[width=\textwidth]{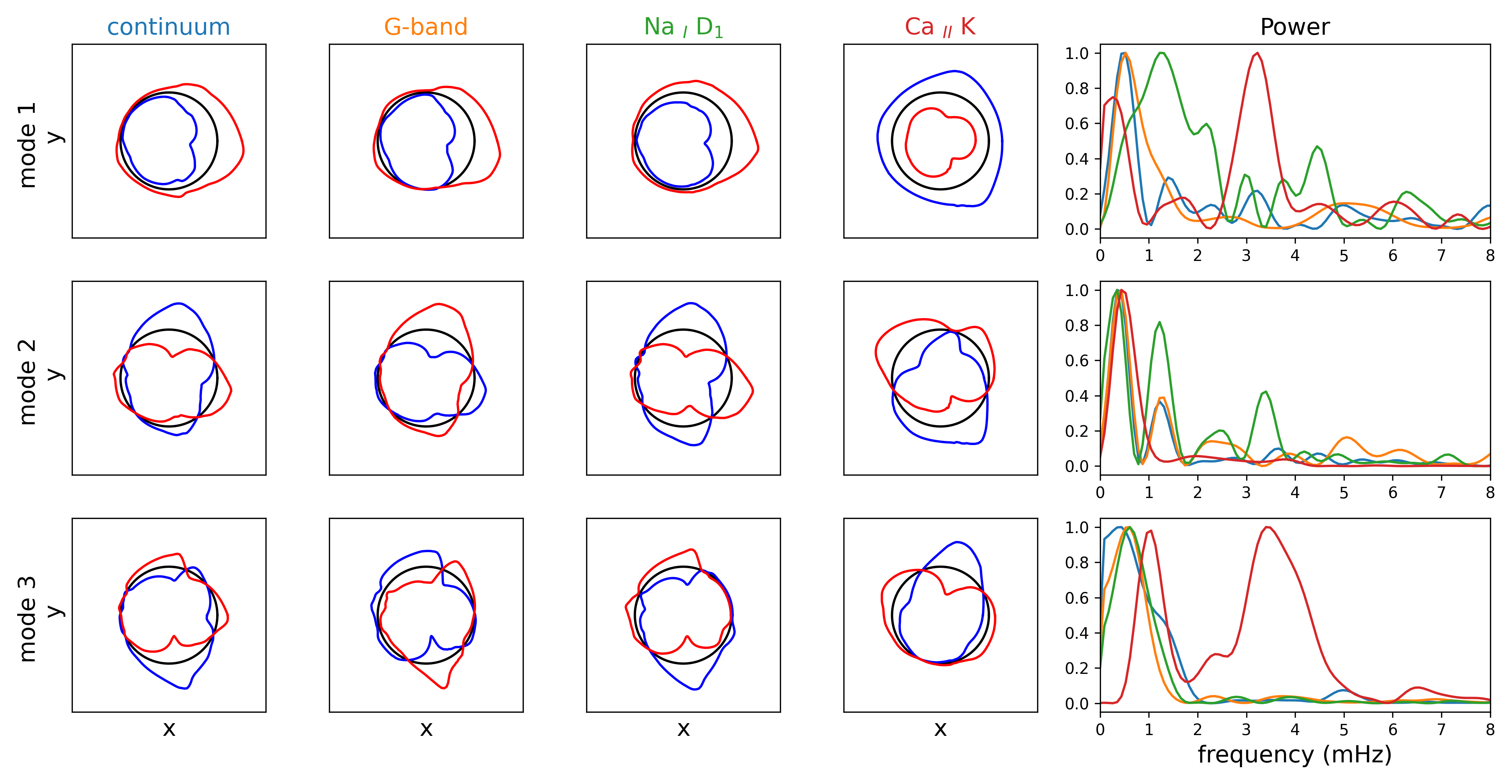}
    \caption{Leading Proper Orthogonal Decomposition (POD) modes of the convexified pore boundary across the four atmospheric heights sampled by the continuum, G-band, Na\,\textsc{i}, and Ca\,\textsc{ii}\,K filters respectively
    Rows correspond to the first three POD modes, ordered by decreasing energy content. 
    In the first four columns, the spatial deformation patterns for each POD mode is shown, with blue and red curves denoting the extrema of the perturbations relative to a circle (black).
    The rightmost column shows the power spectra of the temporal coefficients for the respective order mode, with the different bandpasses denoted by colour (blue:continuum, orange:G-band, green: Na\,\textsc{i}, red: Ca\,\textsc{ii}\,K). 
    Each bandpasses' power spectrum is normalised to its maximum peak, to help compare across heights - this normalisation does not imply that the absolute oscillation powers are directly comparable between passbands.}
    \label{fig:POD_modes}
\end{figure*}

\subsection{POD decomposition of boundary dynamics}
Having established that using the convex hull preserves the physically relevant low-order deformation modes (Section~\ref{subsec:validate}), we now examine the temporal evolution of these modes through POD. 
Each mode is obtained from the convexified boundary and represents an orthogonal component of the boundary’s instantaneous radial displacement about its temporal mean. 
POD yields an ordered set of orthogonal spatial modes, each paired with a time-dependent coefficient.
To reduce frame-to-frame noise, a 
The POD results are displayed in Figure.~\ref{fig:POD_modes}. 

The leading modes map naturally onto familiar solar surface-wave families.
Mode 1 is axisymmetric, corresponding to a sausage-like expansion and contraction of the pore cross-section.
Mode 2 exhibits an elliptical deformation whose major axis rotates in time, characteristic of a kink-like transverse distortion.
Mode 3 shows a more intricate pattern with multiple azimuthal nodes, resembling a low-order fluting disturbance.
Such ordering is common across all the bandpasses analysed, although their relative contribution changes.
Spectral analysis of the temporal coefficients also how the energetics and frequency content of these deformations evolve with height across the continuum, G-band, Na~\textsc{i}, and Ca~\textsc{ii},K bandpasses.

\begin{figure}
    \centering
    \includegraphics[width=\linewidth]{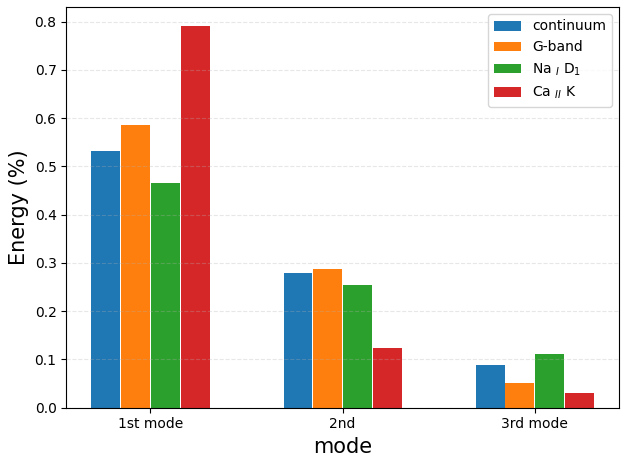}
    \caption{The relative energy of the first three POD modes within the bandpasses.
    }
    \label{fig:mode_energies}
\end{figure}

In Figure.~\ref{fig:mode_energies} we compare the relative energy of the most important POD modes within the bandpasses themselves. 
The first three modes are shown, and when combined are sufficient to recover 80-90\% of the variance. 

\section{Discussion}
\label{sec:discussion}
Mode~1 contains the majority of the boundary-oscillation energy at all heights, with fractional contributions ranging from roughly 50–60\% in the continuum and G-band to nearly 80\% in Ca\,\textsc{ii}\,K. 
This confirms the perceived wisdom: the slow sausage mode is the most important wave mode within a solar pore. 
The corresponding spectra show a consistent low-frequency enhancement. 
In the continuum and G-band, the dominant peak lies near $0.5$\,mHz, while higher layers reveal a systematic shift toward higher frequencies: Na\,\textsc{i} exhibits broad power near $1$–$2$\,mHz, and Ca\,\textsc{ii},K shows a well-defined peak around $3$\,mHz as expected \citep{Miriyala2025}. 
This height-dependent drift is consistent with the behaviour expected of compressive, sausage-like disturbances propagating upward into a progressively rarefied atmosphere. 
The enhanced chromospheric power and the relative narrowing of the Ca,\textsc{ii},K spectrum suggest preferential transmission of these modes into the upper layers, accompanied by damping at lower frequencies. 
Sausage modes therefore dominate the energetic footprint of the pore boundary \citet{Grant2015} and appear to channel energy efficiently into the chromosphere \citet{Grant2022, Gilchrist-Millar2020,Jafarzadeh2024_fluting}.
The close link between flux-tube geometry and oscillatory structure observed by \citet{Stangalini2022_Nature} further underscores the physical relevance of boundary eigenfunctions in interpreting the dominant surface modes.

Mode~2 captures 25–30\% of the boundary variance in the lower layers but drops to only $\sim$12\% in Ca\,\textsc{ii}\,K. 
The spectral signature of the kink mode is remarkably consistent: all four passbands show a strong peak near $0.5$\,mHz, with modest contributions around 1--3\,mHz in the photospheric channels. 
Such low frequencies lie below the expected kink cutoff \citep[we expect a cutoff of some 500--800 seconds i.e. 1--2\,mHz,][figure 7]{Pelouze2023_cutoff}, implying that these displacements are unlikely to represent freely propagating kink waves. 
Instead, the stable $\sim 0.5$\,mHz peak across all heights points to a forced origin, most plausibly granular buffeting of the pore boundary.
The diminishing energy fraction with height reinforces this interpretation: a driven, low-frequency kink response excited in the photosphere would not be expected to propagate efficiently into the chromosphere. 

Mode~3 contributes only a small fraction of the total variance (4--11\%), and its spatial structure is more complex than the first two modes. 
The spectra show a broad low-frequency component in the continuum and G-band, a narrower $\sim 0.8$\,mHz peak in Na\,\textsc{i}, and a combination of a $\sim 1$\,mHz peak with broad 3--5 mHz power in Ca\,\textsc{ii}\,K.
As the MAC analysis has shown, higher order deformations are highly sensitive to small-scale distortions of the pore boundary, and therefore may not be meaningfully preserved when applying a convex hull. Consequently, interpretation of this fluting mode in the boundary must be done with caution.
Furthermore, it has been recently shown that fluting modes are intrinsically fragile in realistic magnetic structures, making them unlikely to remain as coherent oscillations (at least in coronal conditions) due to strong attenuation from nonlinear effects \citep{Soler2025}. 


\section{Conclusions}
\label{sec:conclusions}
Multi-height observations of a solar pore taken by the ROSA instrument at the Dunn Solar Telescope were investigated for boundary dynamics. 
The boundary for each frame in each of the four bandpasses (4170\si{\angstrom} continuum, G-band, Na\,{\sc{i}}, and Ca\,{\sc{ii}}\,K) was extracted using Minimum Error Thresholding, adapting to the instantaneous bimodal intensity distribution in each frame. 

Given the high resolution of our data, combined with the variable fracturing of the pore boundary, we investigated the applicability of the convex hull as a way to recover a stable wave-supporting boundary, making mode analysis feasible.
We computed geometric eigenmodes of both the observed pore shapes and their convex hulls and compare them using the modal assurance criterion.
We show that the convex hull faithfully preserves the fundamental sausage ($m=1$) mode and retains a stable representation of the kink ($m=2$) mode, whereas the higher-order fluting modes ($m\ge3$) are generally no longer reliably captured, unless the original boundary shape is nearly convex. 
This is unsurprising, since small scale irregularities affect the higher order modes far more. 
Furthermore, it is found that the wave modes found from the convex hull becomes less representative of the analytic wave modes as the measured boundary departs too far from a compact geometry in line with expectations.  Such departure from an ideal shape can be quantified as low circularity or high roughness. 
Nevertheless using the convex hull as a surrogate boundary allows techniques such as POD to be applied consistently, and we have shown that the low-order deformation modes of the shape are retained well under convexification. 
This establishes a practical methodology for wave-mode analysis in real pores and provides new constraints on wave behaviour and energy transport in the lower solar atmosphere.

Applying POD to the (convex hull) boundary dynamics reveals clear sausage, kink, and fluting components at all atmospheric heights. 
Two clear trends are shown.
Firstly, the axisymmetric sausage-like mode contributes the largest share of the boundary variance at all heights, dominating the pore dynamics.
Their upward shift in frequency with height and their increasing energetic contribution into Ca\,\textsc{ii}\,K reflect the expected behaviour of upwardly propagating compressive perturbations, travelling through a stratifying and magnetically constrained atmosphere.
These results reinforce previous findings that pores act as efficient waveguides for slow, compressive oscillations \citep{Grant2015,Gilchrist-Millar2020,Guevara-Gomez2023} using complementary methods, 
while extending such analyses beyond just the dominant sausage oscillations by also assessing higher-order boundary modes and mode mixing.
Secondly, the kink-like modal response is weak, largely confined to a persistent low-frequency peak, and therefore is plausibly interpreted as being forced from granular buffeting. 

This study clarifies the geometric regime in which boundary analysis is reliable, and establishes a practical framework using the convex hull for analysing the complex pore geometries captured in high resolution observations. 
This is particularly relevant for modal decomposition techniques such as POD and Dynamic Mode Decomposition (DMD), where physical interpretation often relies on the inferred frequency spectrum. 
While the present work shows that the dominant, low-order geometric modes can be identified robustly under convexification, we note that the extent to which using the convex hull affects the inferred oscillatory frequencies and their errors remains unclear, particularly given that the underlying boundary geometry itself affects the (eigen)spectrum. 
Establishing this connection could form a valuable direction for future work, as would the use of two-dimensional POD to circumvent these limitations by analysing the full spatio-temporal structure of the pore.

\begin{acknowledgements}
TJD and DBJ thank the support of the Leverhulme Trust via the Research Project Grant RPG-2019-371. 
TJD, DBJ, SDTG, and SJ wish to thank the UK Science and Technology Facilities Council (STFC) for the consolidated grants ST/T00021X/1 and ST/X000923/1. 
DBJ and SDTG also acknowledge funding from the UK Space Agency via the National Space Technology Programme (grant SSc-009).
SJ acknowledges support from the European Research Council under the European Union Horizon 2020 research and innovation program (grant agreement No. 682462) and from the Research Council of Norway through its Centres of Excellence scheme (project No. 262622). 
LS acknowledges the STFC for support from grant ST/X001008/1.
SSAS is grateful to the Science and Technology Facilities Council (STFC) grants ST/V000977/1, ST/Y001532/1, UKRI1165; to The Royal Society, IEC/R3/233017 - International Exchanges 2023 Cost Share (NSTC), collaboration with Taiwan. SSAS  would like to thank the International Space Science Institute (ISSI) in Bern, Switzerland, for the hospitality provided to the members of the team on `Opening new avenues in identifying coherent structures and transport barriers in the magnetised solar plasma’.
Finally, we wish to acknowledge scientific discussions with the Waves in the Lower Solar Atmosphere (WaLSA; \href{https://www.WaLSA.team}{https://www.WaLSA.team}) team, which has been supported by the Research Council of Norway (project no. 262622), The Royal Society \citep[award no. Hooke18b/SCTM;][]{2021RSPTA.37900169J}, and the International Space Science Institute (ISSI Team~502). 
\end{acknowledgements}

\bibliography{PODbib}{}
\bibliographystyle{aasjournal}

\begin{appendix}
\section{Additional MAC matrices}
\label{appendix:MAC}
To complement the representative G-band example shown in the main text, we present additional MAC matrices for three further pore boundaries. 
These particular boundaries were chosen because they span the observed range of geometric metrics in Table~\ref{tab:shape_metrics}.  
These examples and their resulting MAC matrices illustrate how modal correspondence behaves as the boundary transitions from compact and mildly corrugated to highly indented and fragmented. 
The Ca\,\textsc{ii}\,K example lies in the regime where the convex hull preserves the lowest-order modes, whereas the continuum and Na\,\textsc{i} examples showcase the degradation of modal correspondence when the boundary is too irregular to support coherent large-scale eigenmodes. 

Furthermore, using the convex hull also produces a systematic reduction in the geometric eigenfrequencies. 
In the Ca\,\textsc{ii}\,K pore example, the shifts are modest ($\approx$ 9–15\%), whereas the continuum example shows larger decreases ($\approx$19–31\%). 
The Na\,\textsc{i} pore example, whose boundary is the most irregular, exhibits the largest changes ($\approx$36–51\%), consistent with the breakdown of modal correspondence seen in its MAC matrix.
We reiterate that the geometric eigenfrequencies shifts introduced by convexification do not affect the POD temporal spectra, which are derived directly from the observed boundary dynamics.


\begin{figure}[h!]
\centering
\includegraphics[width=0.92\linewidth]{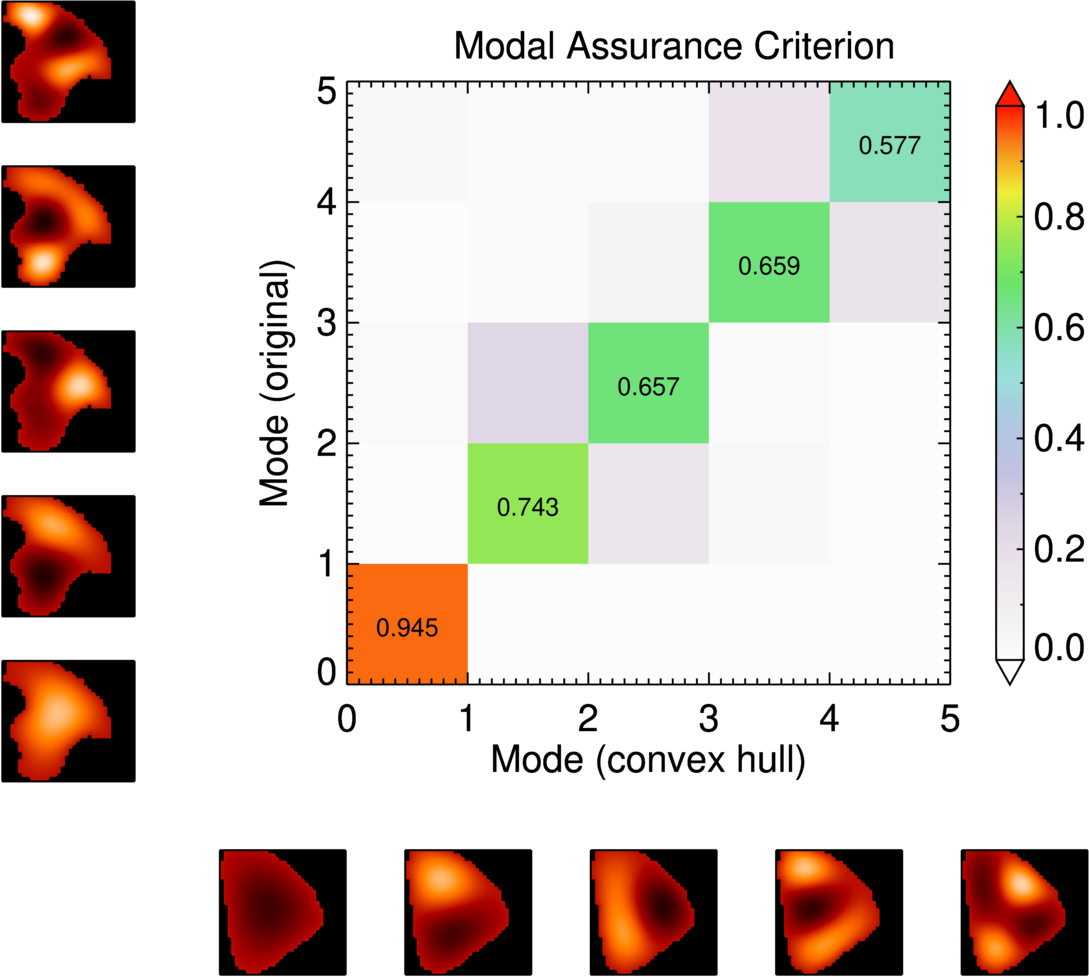}
\caption{MAC matrix for the Ca\,\textsc{ii}\,K pore boundary, showing preservation of the lowest-order modes under convexification.}
\label{fig:MAC_appendix_cak}
\end{figure}

\begin{figure}
\centering
\includegraphics[width=0.92\linewidth]{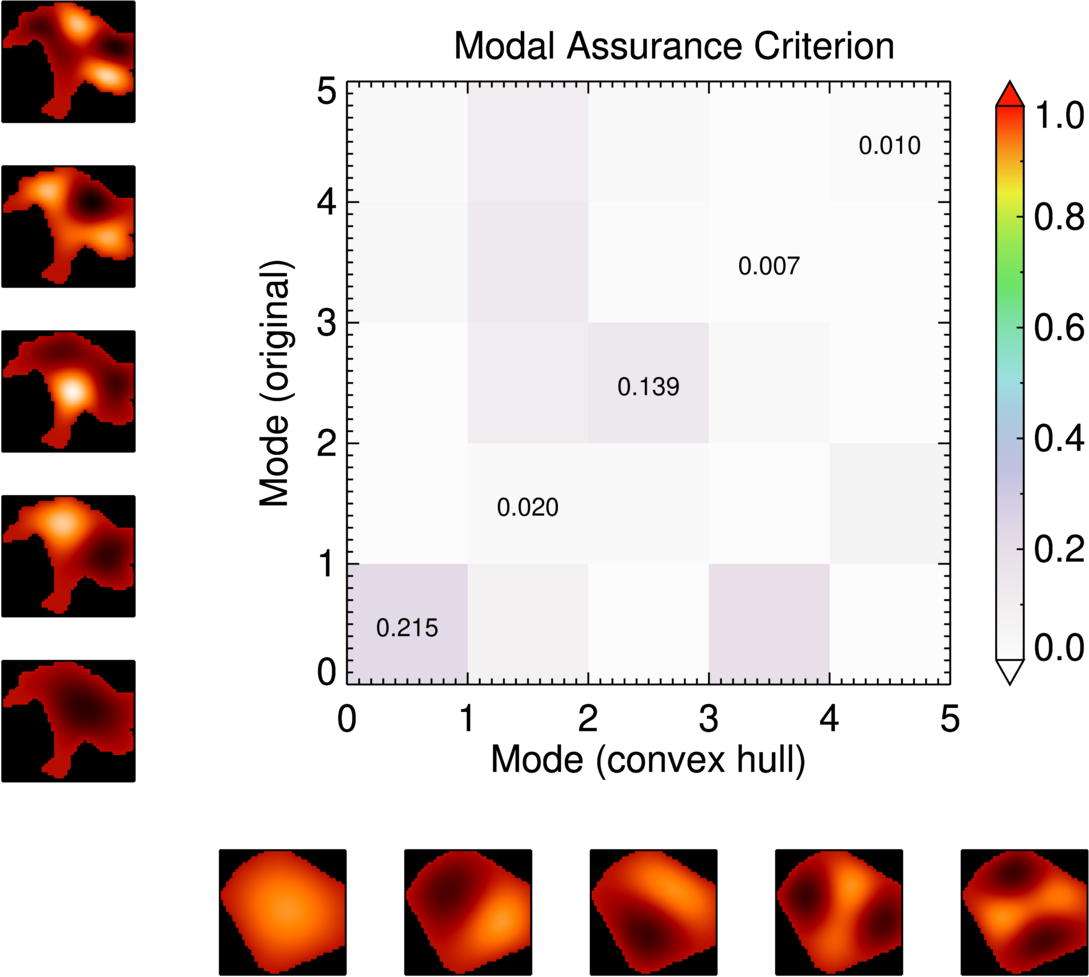}
\caption{MAC matrix for the continuum pore boundary, showing degraded modal correspondence owing to boundary irregularity.}
\label{fig:MAC_appendix_cont}
\end{figure}

\begin{figure}
\centering
\includegraphics[width=0.92\linewidth]{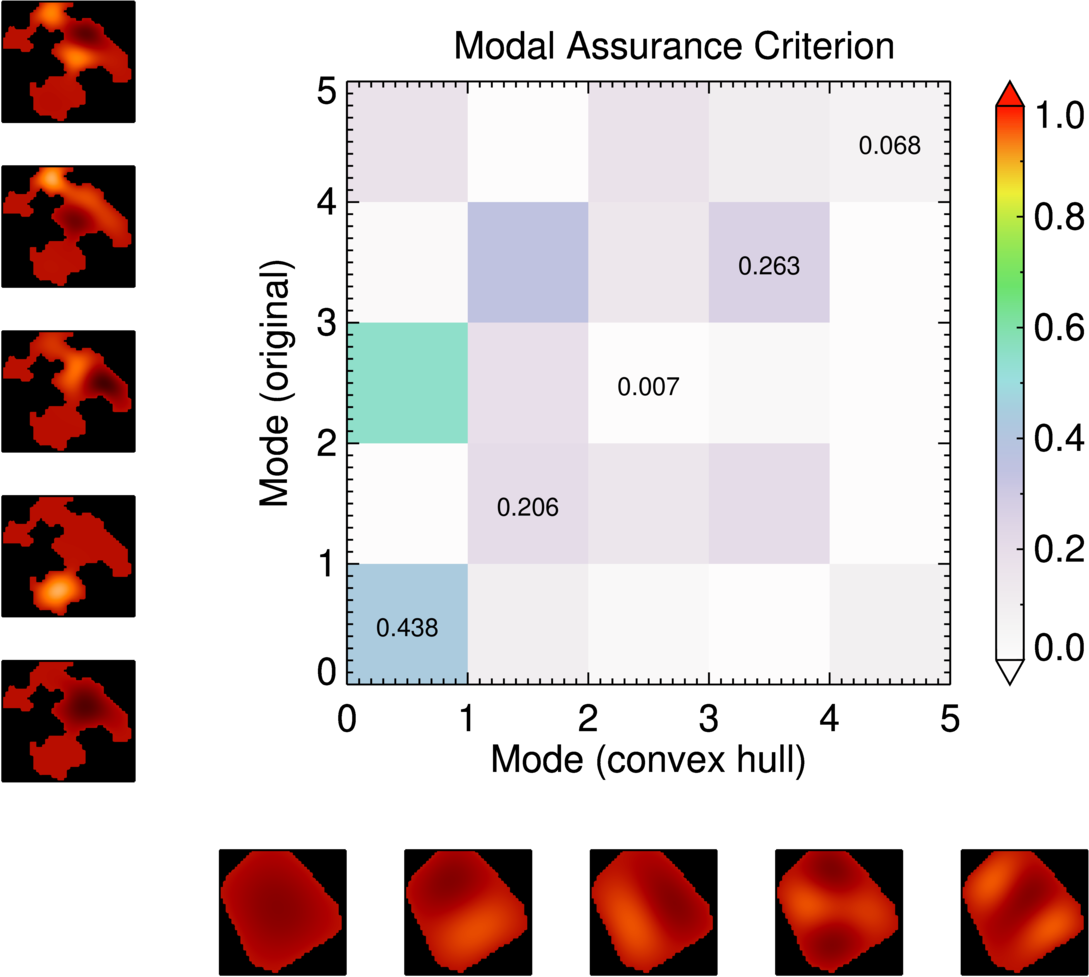}
\caption{MAC matrix for the Na\,\textsc{i} pore boundary, showing strongly fragmented modal correspondence under convexification.}
\label{fig:MAC_appendix_na}
\end{figure}

\end{appendix}

\end{document}